\documentstyle[aps,prl,twocolumn,epsf]{revtex}
\begin{document}
\draft
\flushbottom
\twocolumn[
\hsize\textwidth\columnwidth\hsize\csname @twocolumnfalse\endcsname

\title{Effect of planar impurities on the superfluid density of striped 
cuprates}
\author{A.~H.~Castro~Neto $^1$ and A.~V.~Balatsky $^2$}

\address{$^1$ Department of Physics,
University of California,
Riverside, CA, 92521 \\
$^2$ Los Alamos National Laboratory, Los Alamos, NM 87545}
\date{\today}
\maketitle
\tightenlines
\widetext
\advance\leftskip by 57pt
\advance\rightskip by 57pt

\begin{abstract}
We propose a mechanism for superconductivity suppression in stripe-correlated 
cuprates based on the pinning of the stripes by
impurities, such as Zn. The suppression of  superfluid density occurs in the 
vicinity of the impurity due to the low dimensional character of superfluid carriers 
on 
the stripes. Simple geometric argument about the planar fraction of the 
carriers, 
affected by stripe pinning, leads to  prediction for impurity critical 
concentration $z_c \sim T^2_c$ and for the linear $T_c$ suppression by Zn 
doping. 
We show that our results reproduce the Uemura 
relation in the pure system and we predict the behavior
of the superfluidy density as a function of Zn concentration. 
We compare our results with available data. 

\end{abstract}

]
\narrowtext

One the most striking properties of high temperature superconductors
(HTC) is the sensitivity of the critical temperature, $T_c$, and
the planar superfluid density, $\sigma_s$, to planar impurities which are
introduced with the substitution of the Cu atoms in the CuO$_2$ planes. 
  $T_c$ is suppressed with a few percent
of doping almost independently of the magnetic nature of the impurity 
\cite{xiao}. In particular it has been shown experimentally
that the HTC undergo a superconductor to 
insulator transition due to Zn doping
\cite{fuku}. The fast depression of $\sigma_s$ with Zn has been studied 
extensively by NMR, $\mu$SR \cite{msr,nachumi} and infrared studies
\cite{basov}
and a heated debate has been developed on the interpretation of
the experimental data \cite{debate}. It has also been established
that this suppression is more robust in the underdoped compounds
\cite{will}. The reduction of $T_c$ has   been assigned to
formation of magnetic defects \cite{maha}, the electron scattering
by disorder in the presence of a d-wave
order parameter \cite{dwave,momo}, unitary scattering \cite{unitary}
and local variation of the superconducting gap\cite{cheese}. 

In this paper we would like to introduce a new scenario for the destruction
of superfluid density which is based on the stripe picture of HTC. We propose 
that Zn sites pin (or slow) stripes, so that the stripes form a quasistatic 
mesh.  
The planar superfluid carrier density is suppressed in the vicinity of the Zn 
sites 
on the scale of interstripe distance $\ell$. This effect resembles the 
``swiss cheese''-like state with Zn sites creating voids in the 
superconducting mesh \cite{nachumi}. 
Within these approximations, assuming that all the superfluid carriers are 
localized on the stripes, we find the rate of $T_c$ suppression and critical Zn 
concentration required to completely suppress superconductivity in the 
simplified 
stripe model. The theoretical results fit the experimental data for underdoped 
La$_{2-x}$Sr$_x$CuO$_{4}$, see Figs. 2,3.

Striped phases of holes have been observe experimentally in 
the superconductor 
La$_{1.6-x}$Nd$_{0.4}$Sr$_x$CuO$_{4}$
\cite{tranquada} and 
La$_{2-x}$Sr$_x$NiO$_{4+y}$ \cite{nickel} which is an antiferromagnetic
insulator. Moreover, stripe formation provides a simple explanation
for the observed magnetic incommensurability in 
La$_{2-x}$Sr$_x$CuO$_{4}$ \cite{LSCO}, YBa$_2$Cu$_3$O$_{7-\delta}$
and Bi$_2$Sr$_2$CaCu$_2$O$_{8-x}$ \cite{YBCO}. The magnetic
incommensurability appears in neutron scattering by the splitting
of the commensurate peak at ${\bf Q}=(\pi/a,\pi/a)$ ($a \approx 3.8 \AA$
is the lattice spacing) by a quantity $\delta$ which in the stripe
picture is inversely proportional to the
interstripe distance, $\ell$ \cite{zeit}.
In a recent
paper one of us (A.~H.~C.~N.), proposed an explanation for 
the observed relationship
between $T_c$ and $\delta$, namely, $T_c \propto \delta$, 
as seen in neutron scattering \cite{yamada} and the
suppression of $T_c$ with Zn doping towards a superconductor-insulator
transition \cite{fuku} in terms of the pinning of stripes by Zn
\cite{ahcn}. As shown in \cite{ahcn} this problem is then formally equivalent
to a set of coupled shunted Josephson junctions which undergo a 
Kosterlitz-Thouless
transition (KT) at $T_c$ where  (we use units such that
$\hbar=k_B=1$)
\begin{eqnarray}
T_c(x) = \frac{c}{\kappa_c \ell(x)} = \frac{569}{\ell(x)} \propto \delta
\label{tc}
\end{eqnarray}
where $\ell$ is given in $\AA$, 
$c = 0.2 $ eV $\AA$ is the velocity of propagation of the charge
degrees of freedom and 
$\kappa_c = 0.35$ - is the KT coupling constant. In what follows we
will consider the physics of underdoped compounds like 
La$_{2-x}$Sr$_x$Cu$_{1-z}$Zn$_z$O$_{4}$ where the stripe
formation is a possibility. 

In the case of La$_{1.6-x}$Nd$_{0.4}$Sr$_x$CuO$_{4}$ 
neutron scattering
indicates that the stripes organize themselves into a
set of parallel chains on the CuO$_2$ planes at distance
$\ell$ from each other. This set
of chains rotate $\pi/2$ from plane to plane \cite{tranquada}.
The formation of these structures is associated with the
competition between energy scales generated by the
antiferromagnet and the long range Coulomb repulsion \cite{steve}.

In our approach below we will assume that Zn sites act as a strongly repulsive 
sites which pin stripes.  It is a natural assumption since we know that the 
Zn$^{2+}$ in CuO$_2$  plane will have a closed d shell (d$^{10}$) and any change 
of electronic density on Zn sites, brought upon motion of stripes, will 
require high energy excitations. Hence our assumption that Zn will pin stripes 
and slow their fluctuation in the plane. One can think of Zn sites as an
obstacle which prevents stripes from fluctuating across in the plane. This 
assumption has to be contrasted with the effect of Sr$^-$ ions. Sr sites are 
well separated from the CuO$_2$ planes and  the Coulomb interaction with the 
charges in the plane is weak. That is why we believe the main effect of Sr$^-$ 
is doping of the carriers in the plane. The pinning of stripes by Sr is weak 
if any.

\begin{figure}
\epsfysize7 cm
\hspace{1cm}
\epsfbox{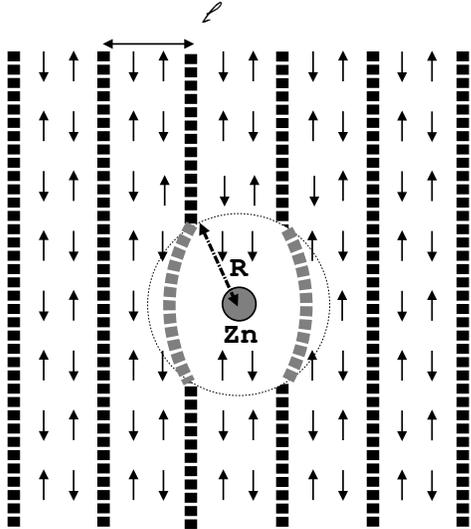}
\caption{Effect of Zn on the stripe grid. Dark dashed lines mark
the regions of stripes with finite superfluid density.
The circle shows the region where superfluid density is suppressed.}
\label{lat}
\end{figure} 

Neutron scattering experiments indicate that the distance
between the stripes is not affected by Zn doping \cite{yamada}.
It implies that the distortions of the stripe lattice, if
exist, have to be local. Because of the repulsion between Zn and
the stripes we assume, for simplicity, that the Zn atoms are
located half distance between stripes as shown in 
Fig.\ref{lat}. 
This is equivalent to the formation of
a {\it topological defect}. A similar effect would happen
when two fluctuating stripes touch each other \cite{zaanen}. 
Since it has been demonstrated experimentally, as in the case
of La$_{1.6-x}$Nd$_{0.4}$Sr$_x$CuO$_{4}$, that
when stripes are pinned by lattice distortions, superconductivity
is suppressed it is natural to assume that superconductivity is
associated with stripe fluctuations. In the presence of an
impurity like Zn these fluctuations are suppressed close to
the Zn site. This suppression extends over a circle of 
radius $R$ around the Zn position which sets the range
of interaction between the Zn and the holes.
This distance $R$ is set by the various energies scales in
the problem including the string tension of the stripe and
therefore should scale with the interstripe distance $\ell$.
From Fig.\ref{lat} it is easy to see that the linear region
of the stripes that is affected is $\sqrt{R^2-\ell^2/4}$. Thus,
we parametrize $R = \gamma \ell/2$ with $1<\gamma$. Here $\gamma$
is a doping independent 
phenomenological parameter which depends on the disorder and
energetic details. 
From the neutron scattering experiments \cite{yamada} we know
that $\delta \propto 1/\ell \to 0$ when $x \approx 0.04$ at the
metal-insulator transition implying that $R \to \infty$ in the
insulating phase. This indicates that the effect of Zn is stronger
in the insulating limit. In the underdoped regime ($0.04 < x < 
0.12$) the data indicates $\delta \propto 1/\ell \propto x$ and therefore
$R \propto 1/x$. In the optimally doped regime ($ 0.12 < x < 0.18$)
it is observed that 
$\delta \propto 1/\ell \to $ constant. From that one concludes that
$R \to $ constant
and a crossover to a Fermi liquid regime where the density is uniform.

Assuming that the Zn atoms suppress superfluidity inside 
the circle of Fig.\ref{lat} we find the following results:
{\it i}) in the pure compound the superfluid density is given by
\begin{eqnarray}
\sigma_s(x,0) \approx \frac{n}{569} T_c(x,0)
\label{uemura1}
\end{eqnarray}
which is the Uemura relation \cite{uemura} ($n$ is the linear 
density of electrons on the stripes); 
{\it ii}) we predict that in the zero temperature limit
\begin{eqnarray}
\frac{\sigma_s(x,z)}{\sigma_s(x,0)} =  1 - \frac{z}{z_c(x)}  \, ;
\label{final1}
\end{eqnarray} 
where $z_c(x)$ is the critical Zn concentration for which
the superfluid density vanishes which is given by 
\begin{eqnarray}
z_c(x) = \left(\frac{a(\AA)}{805}\right)^2 
\frac{T_c^2(x,0)}{\sqrt{\gamma^2-1}} 
\, ;
\label{zc1}
\end{eqnarray}
{\it iii}) using the fact that the KT parameter $\kappa_c$ has to
change with Zn content as
\begin{eqnarray}
\frac{\kappa_c(z)}{\kappa_c(0)} = 1+\alpha(x) z
\label{kappa}
\end{eqnarray}
where $\alpha(x)$ is a constant which depends on the geometry of the mesh.
We obtain from (\ref{tc}) and (\ref{kappa}), in the {\it dilute} limit,
\begin{eqnarray}
\frac{T_c(x,z)}{T_c(x,0)} \approx 1 - \frac{z}{z_c(x)}
\label{tczzc}
\end{eqnarray}
where $\alpha(x) \approx 1/z_c(x)$. Notice that in the underdoped compounds 
($x<1/8$)
the  neutron scattering data show that $T_c(x,0) \propto x$ \cite{yamada} 
which implies from (\ref{zc1}) that $z_c(x) \propto x^2$ which is a
non-trivial prediction of our model. 

\begin{figure}
\epsfysize=2.5 truein
\epsfxsize=3.5 truein
\hspace{1cm}
\vspace{0.5 cm}
\epsfbox[70 260 560 540]{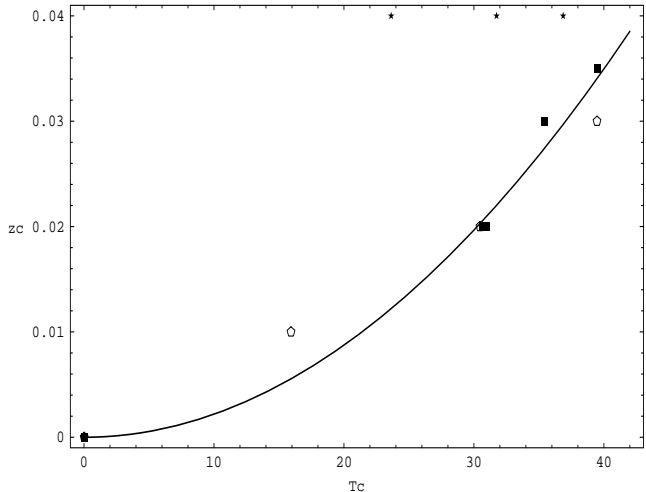}
\caption{$z_c \times T_c$: continuos line: results from (\ref{zc1}); 
filled squares: data from ref \protect\cite{momo}; 
open pentagons: data from ref \protect\cite{will}; 
stars: data from ref \protect\cite{momo} for the overdoped systems.}
\label{zn}
\end{figure}

Some of these results can be compared with the experimental data for
the in-plane penetration depth, $\lambda_{ab}$, 
obtained in $\mu$SR \cite{msr,nachumi}. 
The relaxation rate, $\tau^{-1}$, 
is given by: $\tau^{-1} \propto 1/\lambda_{ab}^2 \propto n_s/m^*$
where $n_s$ is the three dimensional superfluid density and
$m^*$ is the electron effective mass. If $c$ is the distance
between the CuO$_2$ planes then $n_s = \sigma_s/c$. In Fig.\ref{zn}
we show the experimental 
data for $z_c \times T_c$ in La$_{2-x}$Sr$_x$Cu$_{1-z}$Zn$_z$O$_{4}$
\cite{will,momo} and 
the theoretical result with (\ref{zc1}) with $\gamma \approx 1.43$ so that 
$
R \approx 0.71 \ell \, .
$
Close to $x = 1/8$ the critical temperature 
develops a plateau. This plateau is reminiscent of
the suppression of $T_c$ due to structural transitions in 
La$_{1.6-x}$Nd$_{0.4}$Sr$_x$CuO$_{4}$ \cite{tranquada}.
Because of the relationship between $z_c$ and $T_c$ as given in 
(\ref{zc1}) we expect the system to be very sensitive
to Zn content in this region. Thus, the effect of Zn in
suppressing the superfluid density is actually stronger at
the commensurate filling where stripes fluctuate less due
to lattice pining \cite{tranquada}. Indeed, experiments on
the suppression of $T_c$ and on the thermoelectric power
in La$_{2-x}$Sr$_x$Cu$_{1-z}$Zn$_z$O$_{4}$
strongly supports this scenario \cite{koike}.  
The agreement between theory and
experiment is excellent in the underdoped regions where
stripes exist. We observe that in the overdoped
compounds our theory is not applicable since
the system is essentially homogeneous and resembles more
a two dimensional Fermi gas of holes.

\begin{figure}
\epsfysize=2 truein
\epsfxsize=2.7 truein
\hspace{1cm}
\centerline{\epsffile{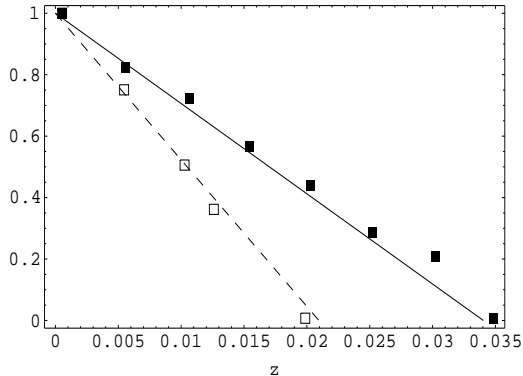}}
\caption{Result of $T_c(x,z)/T_c(x,0)$ from 
(\ref{tczzc}): continuos line: $x=0.16$, dashed line: $x=0.1$.
Experimental data from \protect\cite{momo}: filled squares: $x=0.16$; 
empty squares: $x=0.1$.}
\label{pzc}
\end{figure}

In Fig.\ref{pzc} we show the result of equation (\ref{tczzc}) 
with $z_c(x)$ as given by (\ref{zc1}) for the same parameters
as above for La$_{2-x}$Sr$_x$Cu$_{1-z}$Zn$_z$O$_{4}$ with
$x=0.1$ and $x=0.16$. 
We find a good agreement between our theory and experimental data 
\cite{momo} in the underdoped region with the $T_c$ suppression given by 
Eq.(\ref{tczzc}) with  $z_c(x=0.15) \approx 0.03$ and $z_c(x=0.10) \approx 
0.02$. 

The demonstration of these results is straightforward. 
We assume that there are $M$ stripes in the
CuO$_2$ planes. 
Thus if $L$ is the size of the system then $L=\ell M$
in the direction perpendicular to the stripes.
Each stripe has
a linear density $n$ of electrons so that the Fermi momentum
on each stripe is $k_F = \pi n/2$. The total number of electrons
in stripes on a plane is $M n L = n L^2/\ell$.
In this case the planar density of electrons which participate
on stripe formation is
\begin{eqnarray}
\sigma(x,0) = \frac{n}{\ell(x)} \, .
\label{planar}
\end{eqnarray}
Since the stripes are separated among themselves
by antiferromagnetic regions \cite{zeit} we assume that only
the electrons in stripes can form Cooper pairs. The other
electrons are in a insulating state. This picture seems
to agree with recent numerical calculations of the t-J model
\cite{dmrg} and the application of the stripe picture to the
angle resolved photoemission spectra \cite{paco}. Thus, at
zero temperature and in the absence of planar impurities,
the superfluid density is also given by (\ref{planar}), that is,
$\sigma_s=\sigma$. 
By direct comparison of (\ref{planar}) with (\ref{tc}) we find
\begin{eqnarray}
\sigma_s(x,0) = \frac{n \kappa_c}{c} T_c(x,0)
\label{uem}
\end{eqnarray} 
and using (\ref{tc}) we derive (\ref{uemura1}).
In La$_{1.6-x}$Nd$_{0.4}$Sr$_x$CuO$_{4}$
it is found that $n=1/4$ and it does not seem to change with
doping \cite{tranquada}.

In the presence of impurities the situation is changed because
stripes are one dimensional objects are therefore will be greatly
affected by impurity presence. Neutron scattering data 
indicates that the interstripe distance is not strongly affected
by planar impurities and therefore $\ell=\ell(x)$ only \cite{yamada}. 
As it is shown in Fig.\ref{lat} the number of stripe
electrons which do not participate on the superfluidity per
Zn atom is $4 n \sqrt{R^2-\ell^2/4} = 2 \ell n \sqrt{\gamma^2-1}$. 
Moreover, we 
assume that all the $N_{Zn}$ Zn atoms take part in pinning
the stripes. Thus, the total number of electrons which do not
participate is $N_{Zn} 2 \ell n \sqrt{\gamma^2-1}$ which implies a suppressed
planar density of $N_{Zn} 2 \ell n \sqrt{\gamma^2-1}/L^2$. 
However, we have $z=N_{Zn} (a/L)^2$ so that
the superfluid density which is suppressed is
\begin{eqnarray}
\delta \sigma_s = \frac{2 z \ell n \sqrt{\gamma^2-1}}{a^2} = 
\frac{2 z \ell^2 \sqrt{\gamma^2-1}}{a^2} \sigma(x,0)
\label{cluster}
\end{eqnarray}
where we used (\ref{planar}). 
Finally, the total superfluid density of the system is
given by $\sigma_s(x,z) = \sigma_s(x,0) - \delta \sigma_s$ 
which immediately leads to Eq. (\ref{final1}) where
the critical amount of Zn for which the superfluid
density vanishes is given in (\ref{cluster}) as
\begin{eqnarray}
z_c(x) = \frac{1}{2 \sqrt{\gamma^2-1}} \left(\frac{a}{\ell(x)}\right)^2 \, .
\label{zc}
\end{eqnarray}
Comparing (\ref{zc}) to (\ref{tc}) one finds a non-trivial result
which is given by (\ref{zc1}). 
This result comes out naturally in the stripe model and should not be generally 
expected within homogeneous BCS-like models, where $z_c \propto T_c(x, z=0)$. 
Indeed the superconductivity in Abrikosov-Gor'kov \cite{AG} approach is 
suppressed 
when the scattering rate $1/\tau_s \propto z$ becomes of the order of the 
superconducting gap $\propto T_c$. 

In conclusion, we presented a  ``geometrical'' model for the suppression 
of superfluid density and $T_c$ by impurities within the stripe approach. 
It is based on the assumption that Zn impurities work as local pairbreaker 
and suppresses the superfluid density on stripes in the immediate vicinity of 
Zn. 
We find that our model  fit the data on the underdoped 
La$_{2-x}$Sr$_x$Cu$_{1-z}$Zn$_z$O$_{4}$ compounds with linear $T_c$ suppression, 
Eq.(\ref{tczzc}) and  predicts that $z_c(x) \propto T^2_c(x)$ 
which is consistent with the data.  
 
We are grateful to A.~Bishop, W.~Beyermann, G.~Castilla, 
C.~Hammel, R.~Hefner, D.~MacLaughlin, J.~Sarrao, S.~White, K.~Yamada and 
J.~Zaanen for useful discussions and the Aspen Center for Physics, where 
this work was initiated, for its hospitality.   
A.~H.~C.~N. acknowledges support from the Alfred P.~Sloan foundation and the
partial support provided by the Collaborative University of California - Los 
Alamos (CULAR)  research grant   under the auspices
of the US Department of Energy.

\end{document}